\documentclass{article}
\usepackage{spconf,amsmath,graphicx,float}
\usepackage{booktabs}
\usepackage{multirow}
\usepackage{hyperref} 

\title{Self-supervised learning with speech modulation dropout}
\name{Samik Sadhu$^1$, Hynek Hermansky$^{1,2}$}
\address{$^1$Center for Language and Speech Processing, Johns Hopkins University, USA\\
  $^2$Human Language Technology Center of Excellence, Johns Hopkins University, USA }
  
\begin{document}
%
\maketitle
\begin{abstract}
We show that training a multi-headed self-attention-based deep network to predict deleted, information-dense 2-8 Hz speech modulations over a 1.5-second section of a speech utterance is an effective way to make machines learn to extract speech modulations using time-domain contextual information. Our work exhibits that, once trained on large volumes of unlabelled data, the outputs of the self-attention layers vary in time with a modulation peak at 4 Hz. These pre-trained layers can be used to initialize parts of an Automatic Speech Recognition system to reduce its reliance on labeled speech data greatly. 
\end{abstract}
\section{Introduction}
\label{sec:intro}
Self-supervised learning refers to a machine learning paradigm where supervision is provided by input data itself when conventional labels are not available. In the context of speech signals, a randomly chosen duration of speech features from one utterance is removed from the model input but presented at the output (as supervision) for the model to predict. The machine does so by utilizing information from temporal context in the visible section of the input features from that utterance \cite{oord2018representation,chung2020generative,baevski2020wav2vec,sadhu21_interspeech,hsu2021hubert}. Over the training iterations, the model learns to attend to temporal trends in the feature vectors and gathers patterns which allows it to predict the unseen features. In the process, the model weights are updated to capture contextual information at its output which can be used as compact encoded features for a supervised task such as Automatic Speech Recognition (ASR) in later stages. This makes self-supervised models a popular way to leverage unlabelled data for pre-training a portion of deep network parameters prior to traditional supervised training with labels since it reduces its dependence on large volumes of labeled data while maintaining acceptable error rates.

Rather than removing a portion of the speech from model input, in this work, we investigate if a model is able to learn important speech temporal modulations in 2-8 Hz that are dropped out from a randomly chosen 1.5-second section of a speech utterance at its input. These modulations are presented at the output for the model to predict by attending to available information at its input. We also explore the effects of pre-training the encoder of an End-to-end ASR model with our self-supervised learning technique to study its utility under low-resource supervised ASR training. 

\section{Method}

\subsection{Modulation spectrum}
\label{sec:modulation_spectrum}
The modulation spectrum of speech is given by Fourier analysis of the envelope of the power of speech over a certain duration \cite{658998}. It has been shown that humans communicate primarily through very low modulations with peak information being transmitted around 4 Hz \cite{sadhu2022importance,drullman1994effect,arai1996intelligibility}. Machines are also highly reliant on these modulations for effective speech recognition \cite{kanedera1997importance}.  

\subsection{Computing speech modulation spectrum}
We use complex-FDLP \cite{sadhu22_interspeech} to compute modulations over 1.5-second Hanning windowed speech in 20 frequency sub-bands. Complex-FDLP fits envelopes to the power of a signal over 1.5 seconds. As defined in section \ref{sec:modulation_spectrum}, the Fourier transform of these log envelope fits from complex-FDLP gives the modulation spectrum of speech over these 1.5 seconds. A straightforward way to compute the modulation spectrum is through the complex cepstrum of the complex-FDLP model (see \cite{sadhu22_interspeech} for more details). 

We compute 80 modulation spectrum coefficients that capture modulations below 53 Hz. In the modulation domain, any desired modulations can be removed by forcing their values to be equal to zero. The approximated log speech envelope with desired modulations removed can be computed by taking the Fourier transform of this modified modulation spectrum. 
Speech power envelopes computed for different frequency bands over 1.5-second windows with 50\% overlap can be used to obtain FDLP-spectrograms \cite{sadhu21b_interspeech}.

\begin{figure*}[t]
    \centering
    \includegraphics[scale=0.27]{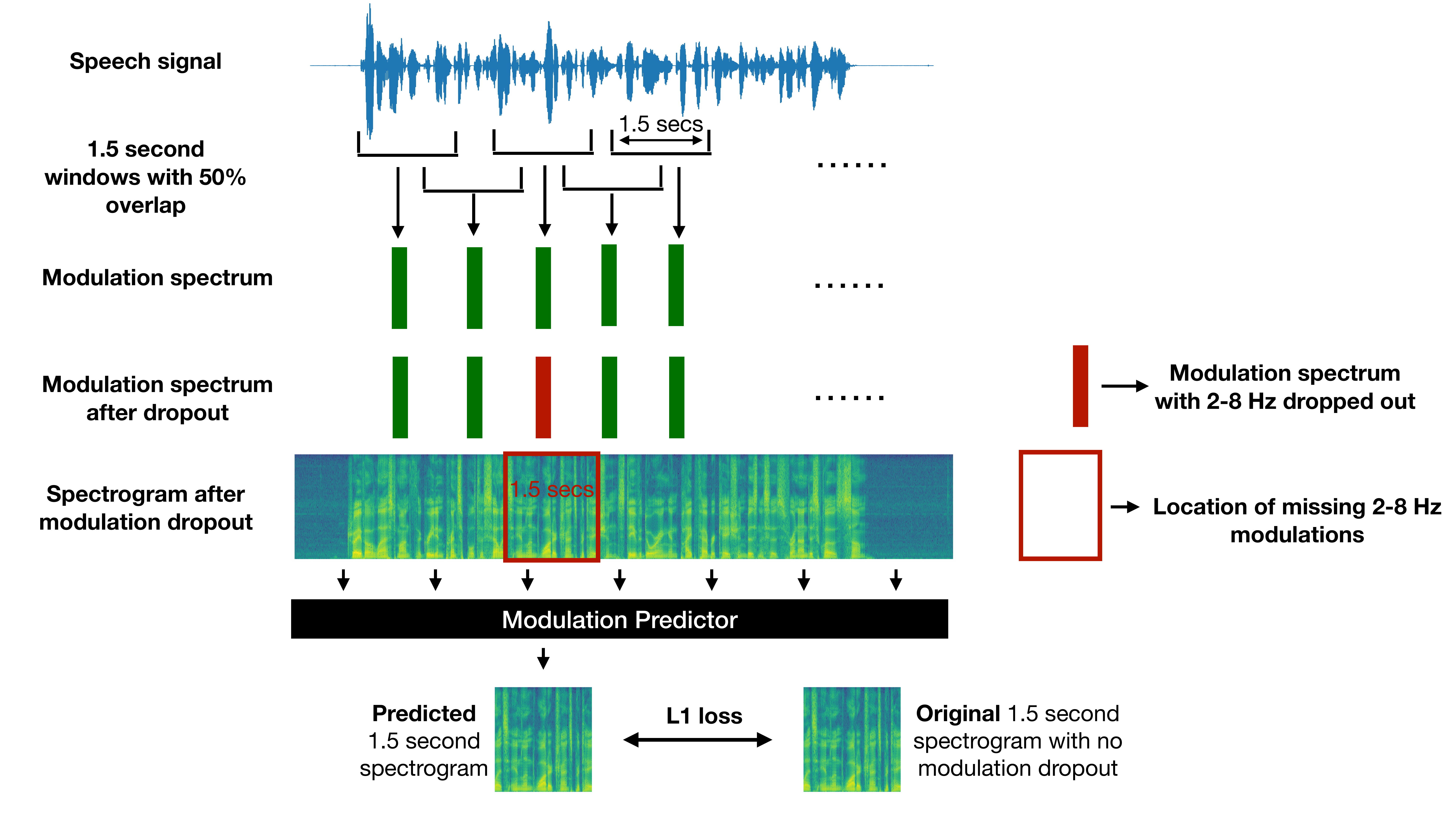}
    \caption{A summary of our self-supervised speech modulation learning is shown in this figure. Modulations are computed for each 1.5-second Hanning windowed speech with a 50 percent overlap. For each utterance, one 1.5-second window is randomly selected for which modulations in the 2-8 Hz range are made zero. The modulations are converted into an FDLP-spectrogram using overlap-add and fed into a modulation predictor which is a neural network tasked with predicting the missing modulations in the 1.5 seconds of speech using contextual information from the whole utterance.}
    \label{fig:modnet}
\end{figure*}
\subsection{Self-supervised modulation predictor}
Figure \ref{fig:modnet} shows our proposed technique for self-supervised modulation learning. Given a speech utterance, we extract 1.5-second Hanning windowed segments with 50\% overlap. Complex-FDLP is used to compute the modulation spectrum in 20 frequency bands in each windowed segment. One modulation spectrum is randomly chosen in each utterance and 2-8 Hz modulations are set to zero. The modulation spectra are converted into FDLP- spectrogram using overlap-add technique \cite{sadhu21b_interspeech} with a 1.5-second section of the spectrogram having 2-8 Hz temporal modulations missing. 

This spectrogram is presented to a deep neural network with multi-headed self-attention layers \cite{vaswani2017attention}. The L1 loss computed between the spectrogram predicted by the deep network and the original FDLP-spectrogram with no modulation dropout over the 1.5 second of interest is used for updating the network parameters during training. 

\section{Experimental Setup}
Our experimental analysis is two-pronged - one had to investigate how the self-attention layers learn to fill out missing modulations and on the other hand explore the effectiveness of a pre-trained modulation predictor used as the encoder in an end-to-end ASR in reducing recognition error rates. 

\subsection{Self-supervised pre-training}
\begin{figure}[h]
    \centering
    \includegraphics[trim={11cm 0cm 11cm 0cm},clip,scale=0.20]{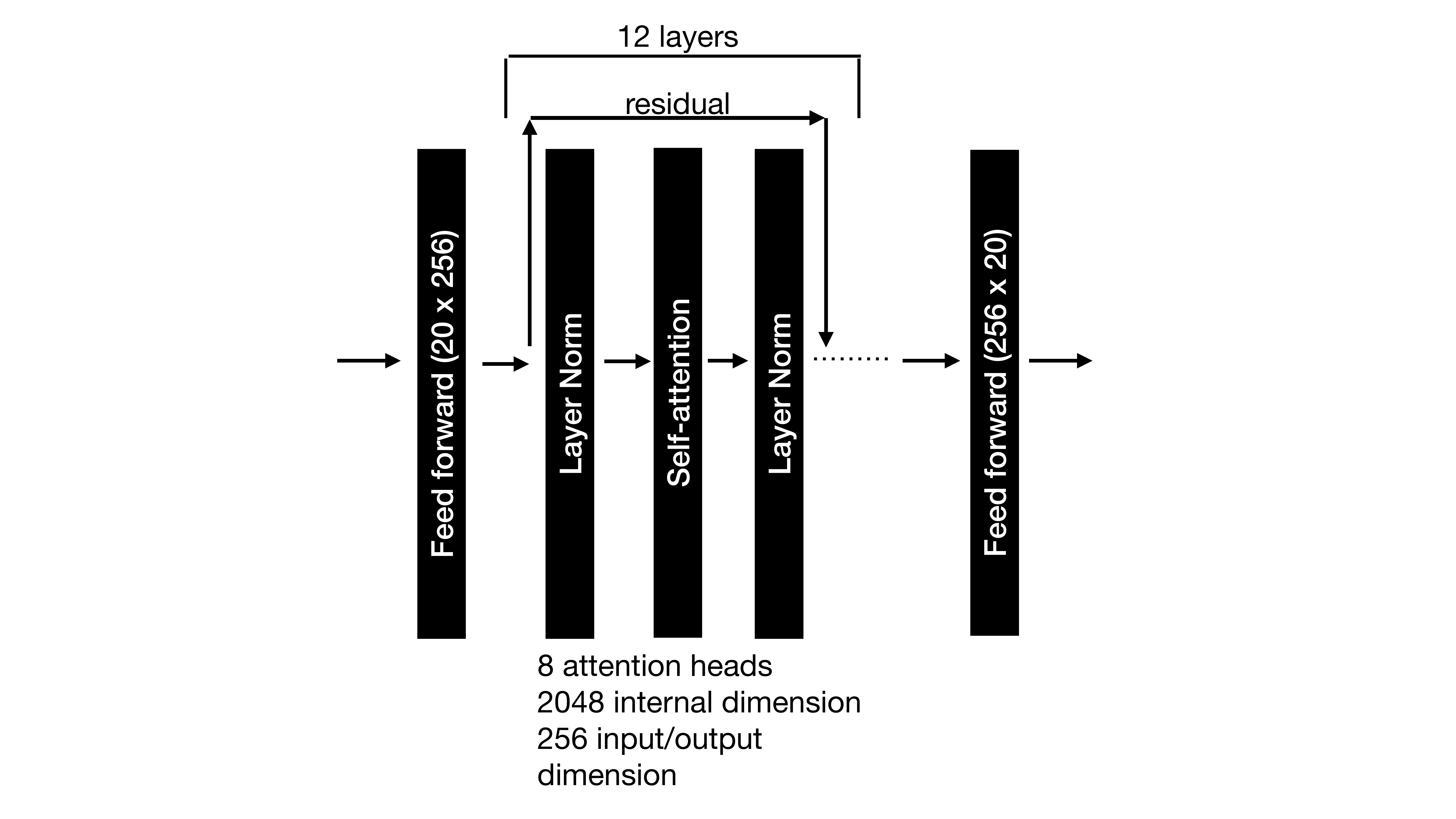}
    \caption{This figure shows the architecture of the modulation predictor neural network. The input 20-dimensional spectrogram is first converted to 256 dimensions followed by 12 self-attention layers with 8 attention heads and 2048 internal dimensions with attached layer norms \cite{ba2016layer} on either side. The 256-dimensional output of the transformer is converted to 20 dimensions at the output to predict a 20-dimensional spectrogram.}
    \label{fig:model_arch}
\end{figure}

We use a multi-headed self-attention-based modulation predictor shown in Figure \ref{fig:model_arch}. Our training data consist of 960 hours of speech from Librispeech data-set \cite{7178964}. The speech training data is augmented 80\% of the time with randomly selected noise from \cite{ko2017study} and SNR is chosen uniformly between 12 to 18 dB. Data augmentation during training for self-supervised learning has been shown to be beneficial towards model robustness \cite{Chen2021WavLM}.

\subsection{Fine-tuning for ASR}
The modulation predictor pre-trained on 960 hours of Librispeech is used as an encoder in a joint CTC-attention end-to-end ASR trained with ESPnet \cite{watanabe18_interspeech}. All the ASR parameters, including the encoder, are then fine-tuned on relatively small 5 or 10 hours of labeled data randomly chosen from the Wall Street Journal speech data set which consists of clean read speech. Alongside the acoustic model, we use a Transformer language model during decoding.

The decoder and the CTC components of all ASRs individually have the same architecture across all experiments. 

\begin{figure*}[h]
    \centering
    \includegraphics[scale=0.25]{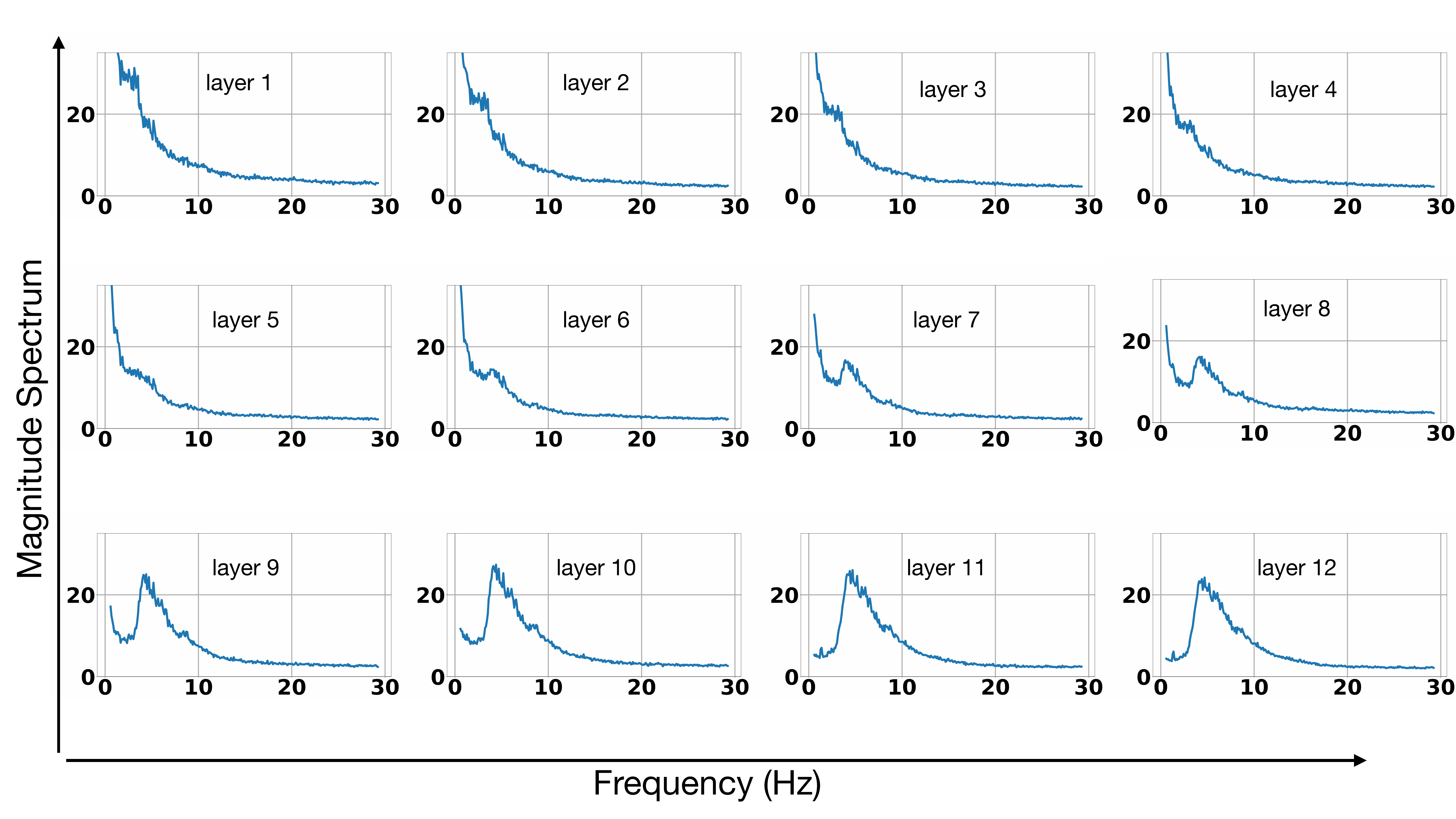}
    \caption{Figure shows temporal modulations captured at the output of different self-attention layers. Deeper into the network, the captured modulations get more refined to have a range of around 2-8 Hz ( which were dropped out at the input to the modulation predictor). Moreover, the model recognizes modulations around 4 Hz to be of maximum importance.}
    \label{fig:layers_of_attention}
\end{figure*}

\section{Results}

\subsection{Analysis of outputs of the self-attention layers}
We analyze the outputs of each of the 12 self-attention layers for 50 randomly chosen utterances from the WSJ test set. Fourier transform is computed along the time axis for each of the dimensions across the 50 utterances. The average magnitude spectra obtained for each of the 12 layers are shown in Figure \ref{fig:layers_of_attention}. This shows the temporal rates of changes or modulations captured by the output of each self-attention layer as we progress across the 12 layers. Interestingly, with the progression of layers, the network learns to capture modulations between 2-8 Hz guided by the training loss. Not only that, but the network also learns to emphasize the modulation peak at 4 Hz which is known to carry most speech information. These results corroborate our previous observations with machine-learned modulation importance values \cite{sadhu2022importance}. 

\subsection{ASR results}
In the conventional way of training a speech recognizer, we train two ASRs on 5 and 10 hours of labeled speech data with the encoder having the same architecture as the modulation predictor where the parameters are randomly initialized. These results are shown in the first column of Table \ref{tab:main_result}. Keeping these ASRs architecturally consistent with our pre-trained ASRs, whose results are shown in the last column in Table \ref{tab:main_result},  allows for a direct comparison of their recognition results. It can be seen that pre-training causes a significant improvement in the recognition error rates as is generally expected.

In future work, we intend to investigate further with varying the amount of training data and training environment to see the effects of our pre-training method on speech recognition results.

\begin{table}[H]
\begin{tabular}{@{}ccc@{}}
\toprule
\multirow{2}{*}{\textbf{\begin{tabular}[c]{@{}c@{}}amount of training data\\ (hours)\end{tabular}}} & \multicolumn{2}{c}{\textbf{WER \%}}                                                   \\ \cmidrule(l){2-3} 
                                                                                                    & \multicolumn{1}{l}{\textbf{no pretraining}} & \multicolumn{1}{l}{\textbf{our method}} \\ \midrule
5                                                                                                   & diverges                                    & 32.3                                    \\
10                                                                                                  & 42.2                                        & 14.3                                    \\ \bottomrule
\end{tabular}
\caption{Word Error Rates (WER \%) are shown on the WSJ test set with three different speech recognizers. Under no pre-training, we train an ASR with a given amount of data with a modulation predictor initialized with regular random initialization. The second column shows the WER with our pre-trained modulation predictor used as the encoder in an ASR and fine-tuned on a given amount of data.}
\label{tab:main_result}
\end{table}

\section{Conclusions}
We show that a deep neural with self-attention layers is capable of learning important speech modulations that have been dropped out of the input to the network over a 1.5-second segment using contextual information. The model learns to replenish the missing modulation gradually over several layers with the output from the ultimate layer capturing temporal modulations with a peak at 4 Hz which is known to carry the most important speech information. After being trained in a self-supervised manner with unlabelled data, this network,  when used as the speech encoder in an end-to-end speech recognition system and fine-tuned with small amounts of labeled data, shows remarkable improvements in recognition accuracy in comparison to a naively trained ASR.

\section{Acknowledgements}
This work was supported  by the Center of Excellence in Human Language Technologies, The Johns Hopkins University, and by a gift from Google Research.
\vfill\pagebreak

\newpage
\bibliographystyle{IEEEbib}
\bibliography{refs}

\begin{thebibliography}{10}

\bibitem{oord2018representation}
Aaron van~den Oord, Yazhe Li, and Oriol Vinyals,
\newblock ``Representation learning with contrastive predictive coding,''
\newblock {\em arXiv preprint arXiv:1807.03748}, 2018.

\bibitem{chung2020generative}
Yu-An Chung and James Glass,
\newblock ``Generative pre-training for speech with autoregressive predictive
  coding,''
\newblock in {\em ICASSP 2020-2020 IEEE International Conference on Acoustics,
  Speech and Signal Processing (ICASSP)}. IEEE, 2020, pp. 3497--3501.

\bibitem{baevski2020wav2vec}
Alexei Baevski, Yuhao Zhou, Abdelrahman Mohamed, and Michael Auli,
\newblock ``wav2vec 2.0: A framework for self-supervised learning of speech
  representations,''
\newblock {\em Advances in Neural Information Processing Systems}, vol. 33, pp.
  12449--12460, 2020.

\bibitem{sadhu21_interspeech}
Samik Sadhu, Di~He, Che-Wei Huang, Sri~Harish Mallidi, Minhua Wu, Ariya
  Rastrow, Andreas Stolcke, Jasha Droppo, and Roland Maas,
\newblock ``{wav2vec-C: A Self-Supervised Model for Speech Representation
  Learning},''
\newblock in {\em Proc. Interspeech 2021}, 2021, pp. 711--715.

\bibitem{hsu2021hubert}
Wei-Ning Hsu, Benjamin Bolte, Yao-Hung~Hubert Tsai, Kushal Lakhotia, Ruslan
  Salakhutdinov, and Abdelrahman Mohamed,
\newblock ``Hubert: Self-supervised speech representation learning by masked
  prediction of hidden units,''
\newblock {\em IEEE/ACM Transactions on Audio, Speech, and Language
  Processing}, vol. 29, pp. 3451--3460, 2021.

\bibitem{658998}
H.~Hermansky,
\newblock ``The modulation spectrum in the automatic recognition of speech,''
\newblock in {\em 1997 IEEE Workshop on Automatic Speech Recognition and
  Understanding Proceedings}, 1997, pp. 140--147.

\bibitem{sadhu2022importance}
Samik Sadhu and Hynek Hermansky,
\newblock ``Importance of different temporal modulations of speech: A tale of
  two perspectives,''
\newblock {\em arXiv preprint arXiv:2204.00065}, 2022.

\bibitem{drullman1994effect}
Rob Drullman, Joost~M Festen, and Reinier Plomp,
\newblock ``Effect of reducing slow temporal modulations on speech reception,''
\newblock {\em The Journal of the Acoustical Society of America}, vol. 95, no.
  5, pp. 2670--2680, 1994.

\bibitem{arai1996intelligibility}
Takayuki Arai, Misha Pavel, Hynek Hermansky, and Carlos Avendano,
\newblock ``Intelligibility of speech with filtered time trajectories of
  spectral envelopes,''
\newblock in {\em Proceeding of Fourth International Conference on Spoken
  Language Processing. ICSLP'96}. IEEE, 1996, vol.~4, pp. 2490--2493.

\bibitem{kanedera1997importance}
Noboru Kanedera, Takayuki Arai, Hynek Hermansky, and Misha Pavel,
\newblock ``On the importance of various modulation frequencies for speech
  recognition,''
\newblock in {\em Fifth European Conference on Speech Communication and
  Technology}, 1997.

\bibitem{sadhu22_interspeech}
Samik Sadhu and Hynek Hermansky,
\newblock ``{Complex Frequency Domain Linear Prediction: A Tool to Compute
  Modulation Spectrum of Speech},''
\newblock in {\em Proc. Interspeech 2022}, 2022, pp. 3208--3212.

\bibitem{sadhu21b_interspeech}
Samik Sadhu and Hynek Hermansky,
\newblock ``{Radically Old Way of Computing Spectra: Applications in End-to-End
  ASR},''
\newblock in {\em Proc. Interspeech 2021}, 2021, pp. 1424--1428.

\bibitem{vaswani2017attention}
Ashish Vaswani, Noam Shazeer, Niki Parmar, Jakob Uszkoreit, Llion Jones,
  Aidan~N Gomez, {\L}ukasz Kaiser, and Illia Polosukhin,
\newblock ``Attention is all you need,''
\newblock {\em Advances in neural information processing systems}, vol. 30,
  2017.

\bibitem{7178964}
Vassil Panayotov, Guoguo Chen, Daniel Povey, and Sanjeev Khudanpur,
\newblock ``Librispeech: An asr corpus based on public domain audio books,''
\newblock in {\em 2015 IEEE International Conference on Acoustics, Speech and
  Signal Processing (ICASSP)}, 2015, pp. 5206--5210.

\bibitem{ko2017study}
Tom Ko, Vijayaditya Peddinti, Daniel Povey, Michael~L Seltzer, and Sanjeev
  Khudanpur,
\newblock ``A study on data augmentation of reverberant speech for robust
  speech recognition,''
\newblock in {\em 2017 IEEE International Conference on Acoustics, Speech and
  Signal Processing (ICASSP)}. IEEE, 2017, pp. 5220--5224.

\bibitem{Chen2021WavLM}
Sanyuan Chen, Chengyi Wang, Zhengyang Chen, Yu~Wu, Shujie Liu, Zhuo Chen, Jinyu
  Li, Naoyuki Kanda, Takuya Yoshioka, Xiong Xiao, Jian Wu, Long Zhou, Shuo Ren,
  Yanmin Qian, Yao Qian, Jian Wu, Michael Zeng, and Furu Wei,
\newblock ``Wavlm: Large-scale self-supervised pre-training for full stack
  speech processing,''
\newblock 2021.

\bibitem{watanabe18_interspeech}
Shinji Watanabe, Takaaki Hori, Shigeki Karita, Tomoki Hayashi, Jiro Nishitoba,
  Yuya Unno, Nelson {Enrique Yalta Soplin}, Jahn Heymann, Matthew Wiesner,
  Nanxin Chen, Adithya Renduchintala, and Tsubasa Ochiai,
\newblock ``{ESPnet: End-to-End Speech Processing Toolkit},''
\newblock in {\em Proc. Interspeech 2018}, 2018, pp. 2207--2211.

\bibitem{ba2016layer}
Jimmy~Lei Ba, Jamie~Ryan Kiros, and Geoffrey~E Hinton,
\newblock ``Layer normalization,''
\newblock {\em arXiv preprint arXiv:1607.06450}, 2016.

\end{thebibliography}

\end{document}